\documentclass[prb,preprint,aps,floatfix]{revtex4-1} %

\usepackage{amsmath} 
\usepackage{amsfonts} 
\usepackage{graphicx} 
\usepackage{float}
\usepackage{hyperref}

\begin{document}


\title{Simulated Gravitational Lensing in the Undergraduate Laboratory}


\author{Daniel FitzGreen}
\email{danfitz@mcmaster.ca}
\affiliation{Department of Physics $\&$ Astronomy, McMaster University, Hamilton, Ontario, Canada L8H 4H3}


\date{\today}

\begin{abstract}

This paper presents a method for fabricating acrylic lenses that simulate the gravitational deflection of light. The fabricated lenses reproduce key features of strong gravitational lensing, while the simulation is further extended to the weak- and microlensing regimes with images analogous to modern astronomical observations. The simulated mass of the lens is measured from the quantitative analysis of the lensed images in each gravitational lensing regime. These independently measured masses are mutually consistent and agree with the values predicted from the machined lens curvature.
\end{abstract}

\maketitle 

\section{Introduction} 

Einstein's theory of general relativity predicts that space and time are distorted by the gravitational field of massive bodies~\cite{Einstein}. Even massless particles, such as photons, have their trajectory altered by the warped space-time of a gravitational field. The image of background objects viewed through the gravitational field of a foreground object is distorted, which is a phenomenon called `gravitational lensing'~\cite{Science}. 

An alternate formulation of gravitational lensing considers Fermat's principle and treats warped space-time as an optical medium with nonuniform index of refraction~\cite{fermat1,Ray,wine1,wine2,wine3,water1}. The most accurate gravitational lens simulations have been done with plastic machined into lenses with the desired curvature~\cite{lens1,lens2,lens3,lens4}. Reference~\citenum{mainlens} contains a thorough description of the construction of plastic lenses that simulate the isothermal cloud- and ideal point-mass distributions. All of these works show images of points or grids deformed into rings and arcs as one would see in an observation of strong gravitational lensing. These works thoroughly explain the calculations and procedures to generate the plastic lenses, but none analyze or measure the resulting images to arrive at a quantitative determination of the simulated mass.

The work presented here demonstrates that acrylic lenses can be used to generate images that are analogous to observations from the three regimes of gravitational lensing. These images can be analyzed to recover the simulated mass of the lens. These measurements are achieved using a single apparatus accessible to most physics undergraduate lab programs and can be completed in a typical three-hour lab session.

This paper introduces a gravitational lensing formalism and the Einstein radius in Section~\ref{sec:form}. Section~\ref{sec:lens} describes the acrylic lenses and additional required components. Section~\ref{sec:regimes} introduces strong, weak, and micro gravitational lensing~\cite{Schneider2006,narayan}, explains the simulated lensing, and presents results. The paper concludes with a discussion of how these experiments can be used in a teaching laboratory, specifically in second- and third-year undergraduate physics labs.  The accompanying supplement contains finer details of image collection and analysis.

\section{Gravitational Lensing Formalism}
\label{sec:form}

\begin{figure}[ht]
\begin{center}
\includegraphics[width=3.5in]{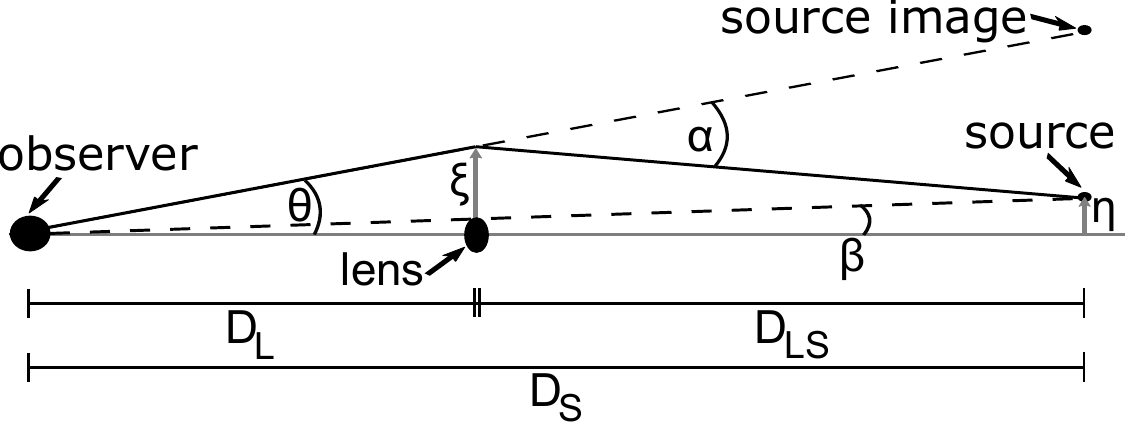}
\caption{Schematic of a gravitational lensing scenario establishing notation.}
\label{fig:GravLensing}
\end{center}
\end{figure}

Figure~\ref{fig:GravLensing} shows an illustration of gravitational lensing. Consider a background source object at a distance $D_S$ from the observer, and a spherical lensing object of mass $M$ at a distance $D_L$ from the observer. The distance between the lensing and source objects is $D_{LS}$. The source object is located at distance $\eta$ and true angle $\beta$ from the center axis (defined by the observer and the lensing object). A ray of light from the source passes by the lens at distance $\xi$ (also called the impact parameter), is deflected by angle $\alpha$, and appears to the observer as if it originated from a source object at an apparent angle $\theta$. 

In the limit of a weak gravitational field ($\Phi/c^2\ll1$, where $\Phi$ is the gravitational potential) of a point-mass lens and ignoring rotation of the lensing object, the trajectory of a photon would be deflected by angle $\alpha$ such that

\begin{equation}
\alpha = \frac{4GM}{c^{2}\xi}.
\label{eq1}
\end{equation}

Presuming the angles are small, then $\theta = \xi/D_{L}$ and the lens equation for this system is

\begin{equation}
\theta^{2}-\beta\theta-\theta_{E}^{2}=0,
\label{quad}
\end{equation}

\noindent where 

\begin{equation}
\theta_{E} =\bigg(\frac{4GM}{c^{2}}\frac{D_{LS}}{D_{S}D_{L}}\bigg)^{1/2} 
\label{eqn:mass}
\end{equation}

\noindent is the so-called Einstein radius, which is typically reported as an angle in units of arcseconds. All angular quantities in this work are converted to radians when used in calculations, though reported in degrees for convenience. Gravitational lensing observations measure the Einstein radius of a lensing system by analyzing changes in the image of the background source object(s). The Einstein radius depends only on mass and distances, so if the distances to the lens and source objects are known and a measurement of the Einstein radius can be made, then astronomers can determine the mass of the lensing object.  

\section{Apparatus}
\label{sec:lens}
\subsection{Acrylic Lenses}

Lenses for this work were designed using the technique from Reference~\citenum{mainlens}, which uses Snell's Law to relate the deflection through the lens ($\alpha$) to its thickness ($T$) as a function of radius ($\xi$):

\begin{equation}
\frac{dT}{d\xi} = -\frac{\alpha(\xi)}{n-1}.
\label{LensEq1}
\end{equation}

\begin{figure}[ht]
\begin{center}
\includegraphics[width=3.5in]{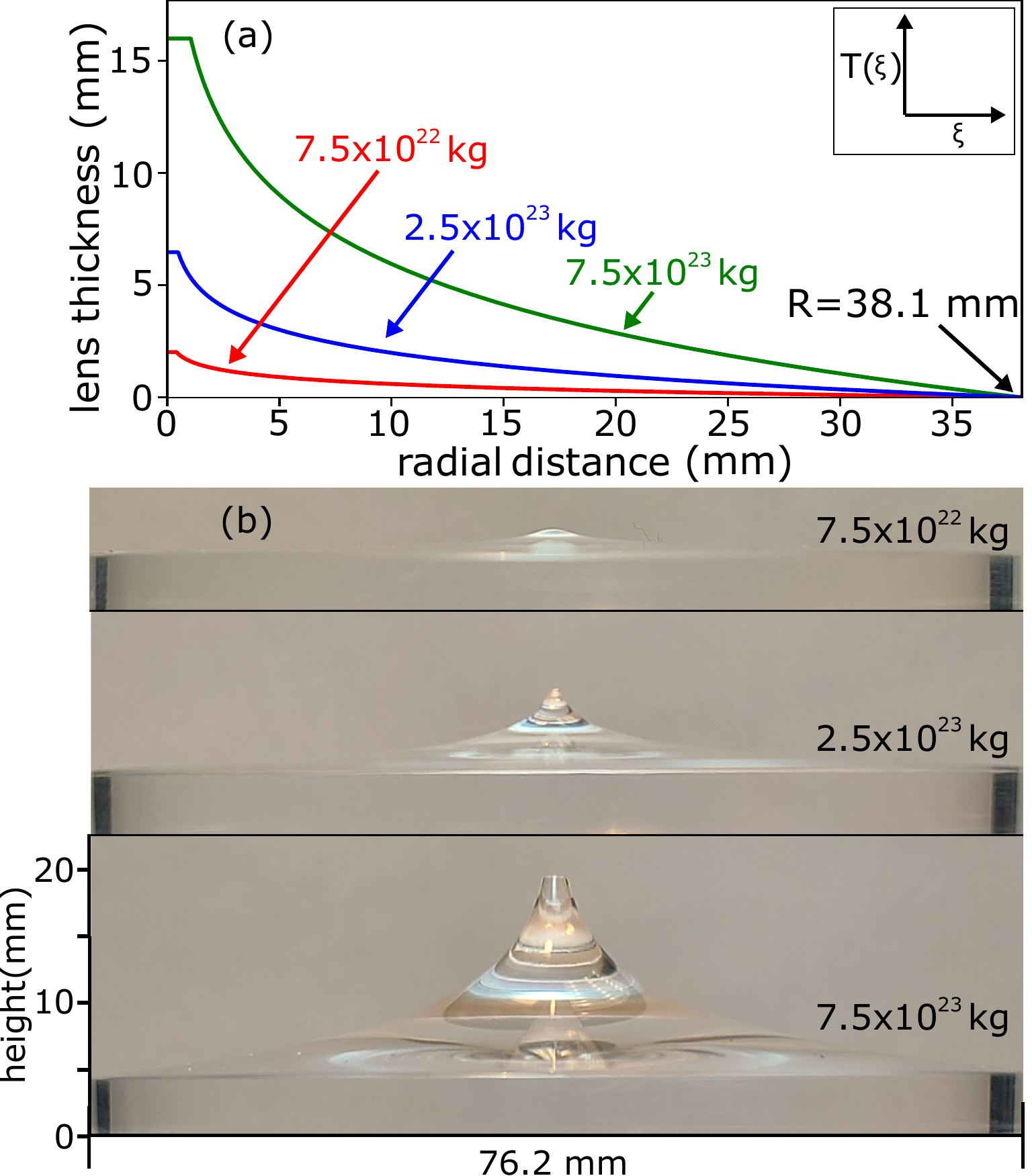} 
\caption{(a) The calculated lens profiles used to simulate point-mass gravitational lenses. The curve has been cut off where the tip of the lens is too thin and brittle, and must be sanded flat. The inset relates the plot to the variables used in Equation~\eqref{eqn:T}. (b) The resulting lenses after they've been cut, sanded, and polished.}
\label{fig:lensthickness}
\end{center}
\end{figure}

Although the acrylic lens is physically thick, the mapping between incident and outgoing rays depends only on the integrated deflection angle, making the thin-lens approximation valid for image reconstruction. Using the deflection angle for a point-mass lensing object from Equation~\eqref{eq1} results in the following lens profile:

\begin{equation}
T(\xi) = T(R) - \frac{4GM}{c^2(n-1)}\ln\bigg[\frac{\xi}{R}\bigg], \xi>0,
\label{eqn:T}
\end{equation}

\noindent where $R$ is the radius of the plastic lens and $n$ is the index of refraction of the lens material ($n=1.5$ for acrylic). Figure~\ref{fig:lensthickness}(a) shows the calculated profiles for the lenses used in this work. These masses generate easily-resolved Einstein rings when the source, lens, and observer are all within 1.0 m (the length of the optical rail). Much larger masses (above $1\times 10^{24}$ kg) would require prohibitively thicker pieces of acrylic, while much smaller masses (below $5\times 10^{22}$ kg) would require too shallow a curvature to be accurately machined and polished.

A Computer Numerical Control (CNC) lathe fabricated these lenses in radial steps of 0.0254 mm (0.001"), then the lenses were sanded with 600-grit sandpaper, buffed with a 3000 grit buffing pad, and polished with headlight lens restorer. Figure~\ref{fig:lensthickness}(b) shows the resulting optically-clear lenses.

\subsection{Additional Components}

\begin{figure}[ht]
\begin{center}
\includegraphics[width=3.5in]{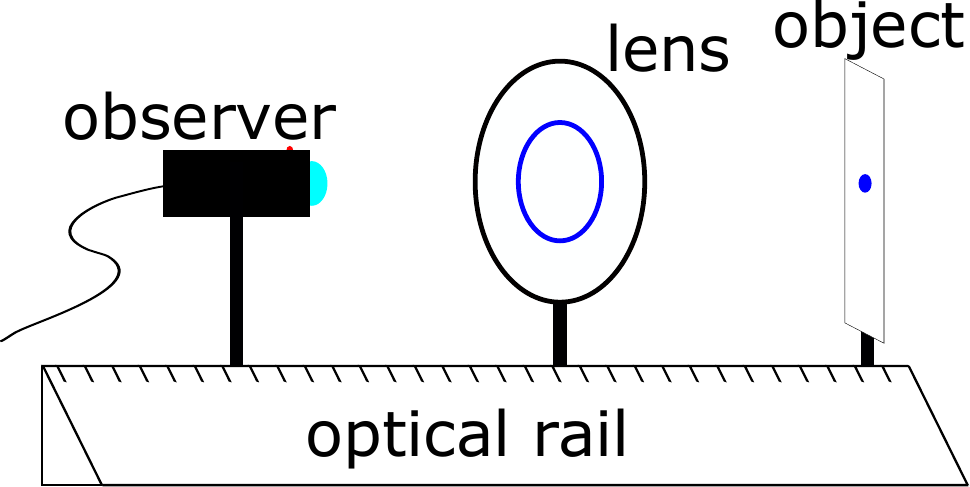}
\caption{A schematic of the simple apparatus used to simulate gravitational lensing in the undergraduate lab. The blue dot (source object) is lensed into a ring by the acrylic lens. An image of the lensed background source object is saved using the webcam (observer). Drawing not to scale.}
\label{fig:apparatus}
\end{center}
\end{figure}

Figure~\ref{fig:apparatus} shows a schematic of the gravitational lensing apparatus used in this work. A webcam (the observer), acrylic lens (the lensing object), and a piece of paper (the source object) are mounted on an optical rail, which is used to measure the distance between components.  A PC running Python with the Open Source Computer Vision Library package (OpenCV) saves frames from the webcam, which are analyzed using ImageJ to determine the simulated mass of the lens. All of these components are readily available in many undergraduate laboratories; however, the simulation of gravitational microlensing additionally requires a linear actuator to translate the lens perpendicular to the optical rail.

\section{Gravitational Lensing Experiments}
\label{sec:regimes}

Acrylic lenses are used to simulate all three regimes of gravitational lensing: strong, weak, and microlensing. These regimes are distinguished by the value of the convergence,$\kappa$, which quantifies the degree to which the lensing mass concentrates light from a background source object, analogous to the focusing power of an optical lens. The convergence for a lensing mass $M$ is the ratio of the surface mass density ($\Sigma(\xi)$), which is the mass per unit surface area from the point of view of the observer, and critical mass density ($\Sigma_{cr}$), which is the minimum mass per unit surface area required to generate a duplicate image of the background source object~\cite{critmass}:

\begin{equation}
\kappa = \frac{\Sigma(\xi)}{\Sigma_{cr}},
\end{equation}

\noindent where 

\begin{equation}
\Sigma(\xi) = \frac{M}{\pi\xi^2}
\end{equation}

\noindent and

\begin{equation}
\Sigma_{cr} = \frac{c^2}{4\pi G}\frac{D_S}{D_LD_{LS}}.
\end{equation}

The strong lensing regime is defined by $\kappa \geq 1$, while weak lensing and microlensing are defined by $\kappa \ll 1$.

\subsection{Strong Gravitational Lensing}

\begin{figure}[ht]
\begin{center}
\includegraphics[width=3.5in]{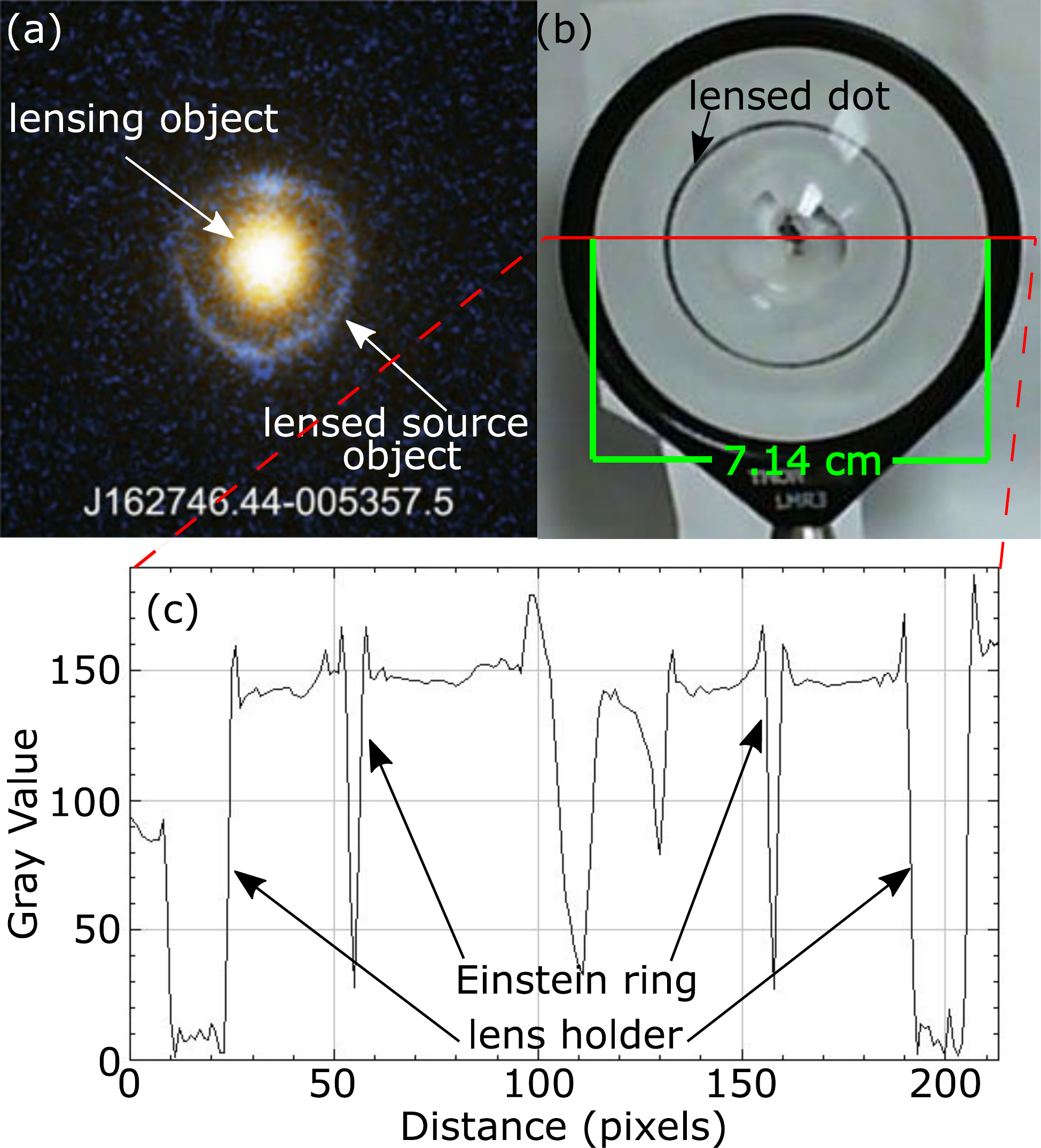}
\caption{ (a) An example of strong gravitational lensing resulting in an Einstein ring. Image credit: NASA, ESA, A. Bolton (Harvard-Smithsonian CfA) and the SLACS Team. (b) An example of strong gravitational lensing of a 5 mm diameter black dot by an acrylic lens simulating a $7.5\times 10^{23}$ kg point-mass gravitational lens. (c) The plot profile in greyscale along the horizontal line across the lens. The known diameter of the lens holder is used to calibrate the radius of the Einstein ring. The dips at the center are the lensed images of the surroundings.}
\label{fig:stronglensing}
\end{center}
\end{figure}

Lensing masses and geometries that meet the strong gravitational lensing criteria generate Einstein rings, arcs, and duplicate images of background source objects. These distortions are the direct result of the quadratic form of Equation~\eqref{quad}, which has two solutions for $\theta$, indicating that a source object will be `mirrored' due to the deflection caused by the lensing object. When the observer, lens, and source object all lie in a straight line ($\beta=0$), then the azimuthal symmetry of the system results in a ring of solutions, and the source object is lensed into a circle around the lensing object called an Einstein ring. Figure~\ref{fig:stronglensing}(a) shows an example of an Einstein ring from the SLACS Survey. If the observer can measure the angular radius of the Einstein ring and distances to the lens and source objects, then the mass of the lensing object can be calculated from Equation~\eqref{eqn:mass}. 

The source object in this simulated gravitational lensing experiment is a 5-mm-diameter black dot on a white sheet of paper. When viewed through the acrylic lens, the image of the dot is focused into a ring around the center of the lens, an example of which is shown in Figure~\ref{fig:stronglensing}(b). The image of the Einstein ring is analyzed using ImageJ. A line profile through the center of the lens (Figure~\ref{fig:stronglensing}(c)) determines the diameter of the lens holder and Einstein ring in units of pixels. The diameter of the Einstein ring is converted from pixels to meters using the known inner diameter of the lens holder (7.14 cm). The Einstein radius is determined from the radius of the Einstein ring and distance to the lens (assuming small angles): $\theta_E\approx r_E/D_L$. In these experiments, $r_E/D_L$ is always less than 0.05. The distances between the observer, lens, and source are measured using the ruler on the optical rail. These measurements are combined using Equation~\eqref{eqn:mass} to determine the simulated mass of the lens.

Table~\ref{tab:strong} shows measurements of the simulated mass of the $7.5 \times 10^{22}$ kg point-mass lens with several values of $ D_S$, $D_L$, and $D_{LS}$. All simulated mass measurements agree with the expected value of the mass based on the machined curvature of the lens. The weighted average of these values is $7.5(2)\times 10^{22}$ kg. 

The largest source of uncertainty in these experiments is the measurement of the radius of the Einstein ring, which is typically $15-100$ pixels with an uncertainty of $\sqrt{2}$ pixels ($\pm1$ pixel for each side of the diameter), thereby limiting the precision of the mass measurement to approximately $10\%$ for the smallest Einstein rings. Systematic errors due to the small angle approximation ($<0.1\%$), thin lens approximation (the radius of curvature is more than ten times the thickness of the lens at the radius of the Einstein ring) as well as the camera distortion ($<1\%$ across the diameter of the lens) are all small enough to be ignored. 

\begin{table}[ht]
\begin{ruledtabular}
\caption{Measurements of strong gravitational lensing using an acrylic lens that simulates a $7.5 \times 10^{22}$ kg point-mass gravitational lens. Values in parenthesis are one standard error.}
\begin{tabular}{ccccc}
$D_S$(m) & $D_L$(m) & $D_{LS}$(m) & $\theta_E$(degrees) & mass (kg) \\
\hline
0.880 & 0.170 & 0.710 & 1.86(5) & 7.5(4)$\times 10^{22}$ \\
0.680 & 0.200 & 0.480 & 1.60(5) & 7.5(5)$\times 10^{22}$ \\
0.880 & 0.280 & 0.600 & 1.33(4) & 7.4(5)$\times 10^{22}$ \\
0.680 & 0.430 & 0.250 & 0.81(4) & 7.9(8)$\times 10^{22}$ \\
0.880 & 0.430 & 0.450 & 0.92(4) & 7.4(7)$\times 10^{22}$ 
\end{tabular}
\label{tab:strong}
\end{ruledtabular}
\end{table}

Acrylic lenses that simulate $7.5\times 10^{22}$ kg, $2.5\times 10^{23}$ kg, and $7.5\times 10^{23}$ kg gravitational lenses generate measurable Einstein rings. The measured radius of the Einstein ring ($r_E$) is in good agreement with the radius where $\kappa=1$ ($r_{\kappa=1}$), which is the criterion used to determine the mass of the lensing object in astronomical observations. The measured mass for each of these lenses (shown in Table~\ref{tab:otherstrong}) is consistent with the expected simulated mass based on the machined curvature of the lens, demonstrating that the same analysis techniques used with images of true strong gravitational lensing can be used to make an accurate determination of the simulated mass of these lenses. 

\begin{table}[ht]
\begin{ruledtabular}
\caption{Measurements of strong gravitational lensing using acrylic lenses that simulate point-mass gravitational lenses. Values in parenthesis are one standard error.}
\begin{tabular}{cccccccc}
$D_S$(m) & $D_L$(m) & $D_{LS}$(m) & $r_{\kappa=1}$(mm) & $r_E$(mm) & $\theta_E$(degrees) &measured mass (kg) & expected mass (kg) \\
\hline
0.880 & 0.430 & 0.450 & 7.0 & 6.9(3) & 0.92(4) & $7.8(7)\times 10^{22}$ & $7.5\times 10^{22}$\\
0.880 & 0.430 & 0.450 & 12.8& 12.7(6) & 1.69(8) & $2.5(2)\times 10^{23}$ & $2.5\times 10^{23}$\\
0.880 & 0.430 & 0.450 & 22.1 & 22.3(6) & 2.97(9)&$7.6(5)\times 10^{23}$ & $7.5\times 10^{23}$
\end{tabular}
\label{tab:otherstrong}
\end{ruledtabular}
\end{table}

\subsection{Weak Gravitational Lensing}

\begin{figure}[ht]
\begin{center}
\includegraphics[width=7in]{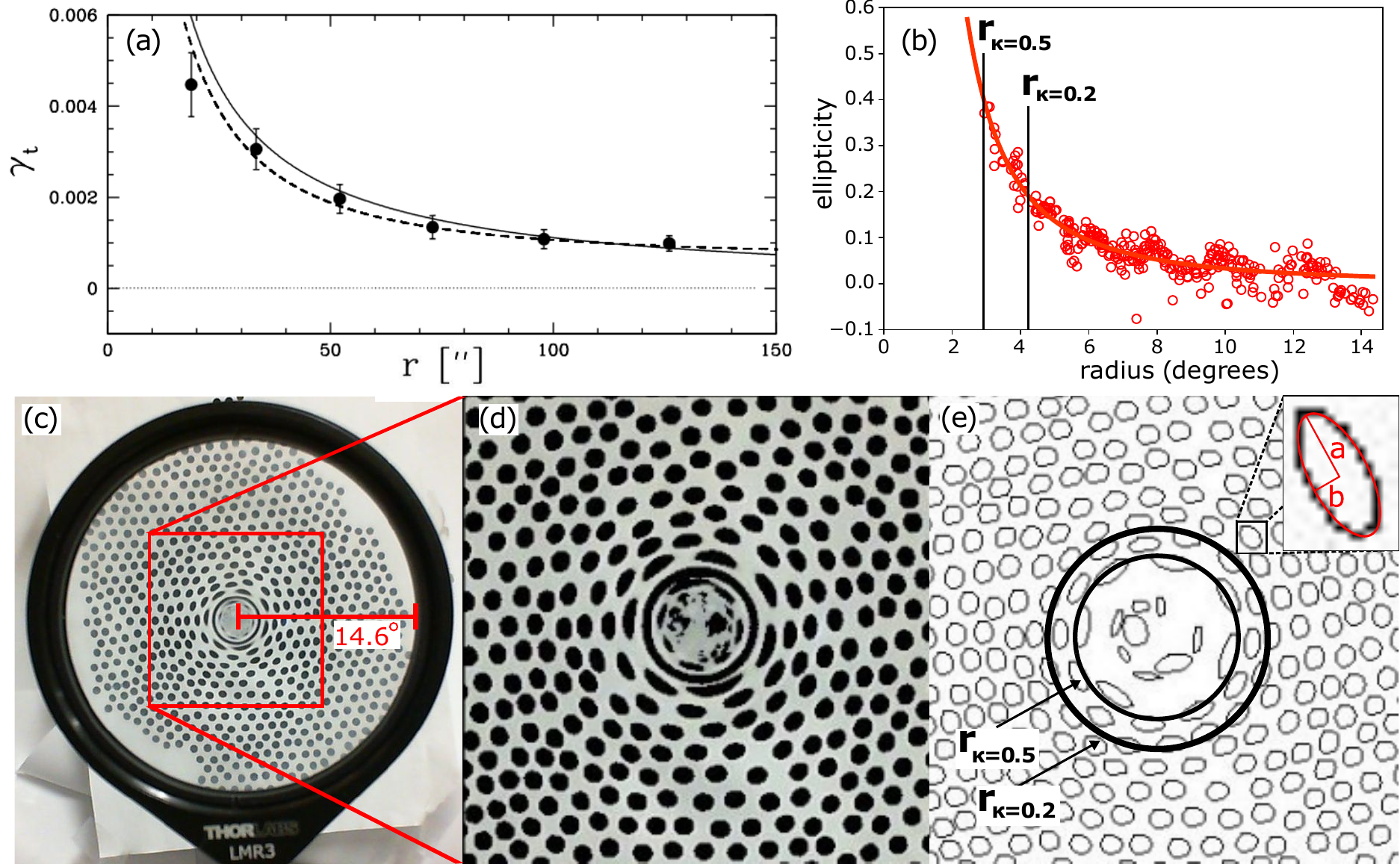}
\caption{(a) An example of galaxy-galaxy weak gravitational lensing~\cite{Parker}\copyright AAS. Reproduced with permission. The data points represent the average tangential shear from many background source objects in a given annulus, which is fit with $\theta_E/\theta$ (solid line) to determine the Einstein radius. (b) An example of simulated weak lensing using an acrylic lens. Ellipticity as a function of radius, with a least-squares fit using Equation~\eqref{eqn:powsq} to find the Einstein radius. (c) Many identical circles viewed through an acrylic lens that simulates a $7.5\times 10^{22}$ kg point-mass gravitational lens. The 7.14 cm diameter lens, 14 cm from the observer, has an angular radius of $14.6^{\circ}$. (d)  ImageJ applies a threshold to the pixel values, transforming them into either black or white.  (e) ImageJ fits each object in the image with an ellipse and determines its position, angle, and major and minor axis values. Ellipses can be excluded based on the value of $\kappa$ at their radial position or if the threshold process blended multiple ellipses together. }
\label{fig:weakdata}
\end{center}
\end{figure}

The weak gravitational lensing regime is defined by $\kappa\ll 1$. Weak lensing observations include many background source objects located far from the center of the lensing object. The lensing object affects the source images weakly enough to leave their apparent positions unchanged, but its tidal gravitational field shears the images by stretching and squishing them. The shear due to weak gravitational lensing should occur tangentially to the angular coordinate in the lens plane, but the measurement is complicated by the source object's inherent ellipticity and orientation, so a statistical approach that measures many background source objects is required. The Einstein radius of the lensing object can be determined by measuring the average tangential shear in the image of the background source objects within an annulus around the lensing object. Figure~\ref{fig:weakdata}(a) shows an example data set from the first galaxy-galaxy weak lensing result, though this fit uses the more realistic singular isothermal sphere density profile: $\gamma(\theta)=\theta_E/\theta$. The shear trend for a point-mass density profile is described by~\cite{Evans}

\begin{equation}
\gamma(\theta) = \left(\frac{\theta_E}{\theta}\right)^2.
\label{eqn:powsq}
\end{equation}

Measurements of $\gamma$ as a function of $\theta$ can be fit with Equation~\eqref{eqn:powsq} to determine $\theta_E$, which is combined with the distances between the observer, lens, and background objects to determine the lensing mass through Equation~\eqref{eqn:mass}.

In this simulation of weak gravitational lensing, the source object is an image of randomly distributed circles with diameters of 5 mm. In the simple case where light from a circular source object passes through the gravitational field of a point-mass lensing object, the image is distorted into an ellipse oriented tangentially to the angular coordinate in the lens plane, as illustrated in Figure~\ref{fig:weakdata}(c). The measured ellipticity ($\epsilon$) is an estimator of shear, where the equality only holds for small shear and intrinsically circular sources:

\begin{equation}
\gamma = \epsilon = \frac{a-b}{a+b},
\end{equation}

\noindent where $a$ is the major axis and $b$ is the minor axis of the ellipse. ImageJ analyzes the image of the lensed circles, applying a colour threshold that separates each pixel into either black or white as in Figure~\ref{fig:weakdata}(d) and fitting the shapes with an ellipse as in Figure~\ref{fig:weakdata}(e). The analysis excludes blended ellipses and retains only those oriented perpendicular to the radial direction and located at a minimum distance from the lens centre, ensuring the convergence remains below a threshold value ($r_{\kappa<0.2}$).  Strictly speaking, $0.2\not\ll1$, so the system does not lie firmly within the weak-lensing regime; nonetheless, the measured ellipticity as a function of radius exhibits the same qualitative trend observed in true weak gravitational lensing.

A least-squares fit with Equation~\eqref{eqn:powsq} determines $\theta_E$, as shown in Figure~\ref{fig:weakdata}(b). Equation~\eqref{eqn:mass} calculates the simulated mass of the lens using the resulting Einstein angle and distances to the lens and source object, as read from the optical rail. Table~\ref{tab:weakdist} shows measured mass values for various distances, yielding a weighted average of $7.52(15)\times 10^{22}$ kg. The choice of threshold value in ImageJ and choice of $r_{\kappa}$ in the ellipse selection introduce systematic uncertainties of $\pm 0.25\times 10^{22}$ kg and $\pm 0.15\times 10^{22}$ kg, respectively. The final result including systematic and statistical uncertainties is $7.5(3)\times 10^{22}$ kg, which is in good agreement with the expected simulated mass of $7.5 \times 10^{22}$ kg.

\begin{table}[ht]
\begin{ruledtabular}
\caption{Measurements of weak gravitational lensing using an acrylic lens that simulates a $7.5 \times 10^{22}$ kg point-mass gravitational lens. The Einstein radius and its uncertainty are determined from the fit as in Figure~\ref{fig:weakdata}(e). Values in parenthesis are one standard error due to the statistical uncertainty from the distance measurements and the fit that determines $\theta_E$.}
\begin{tabular}{ccccc}
$D_S$ (m) & $D_L$ (m) & $D_{LS}$ (m) & $\theta_E$ (degrees) & mass (kg) \\
\hline
0.480 & 0.260 & 0.220 & 1.13(2) & 7.5(3)$\times 10^{22}$\\
0.610 & 0.260 & 0.350 & 1.47(3) & 7.5(3)$\times 10^{22}$\\
0.880 & 0.310 & 0.570 & 1.23(3) & 7.5(4)$\times 10^{22}$\\
0.680 & 0.430 & 0.250 & 0.77(2) & 7.7(5)$\times 10^{22}$\\
0.410 & 0.140 & 0.270 & 1.81(2) & 7.5(3)$\times 10^{22}$
\end{tabular}

\label{tab:weakdist}
\end{ruledtabular}
\end{table}

\begin{figure}[ht]
\begin{center}
\includegraphics[width=3.5in]{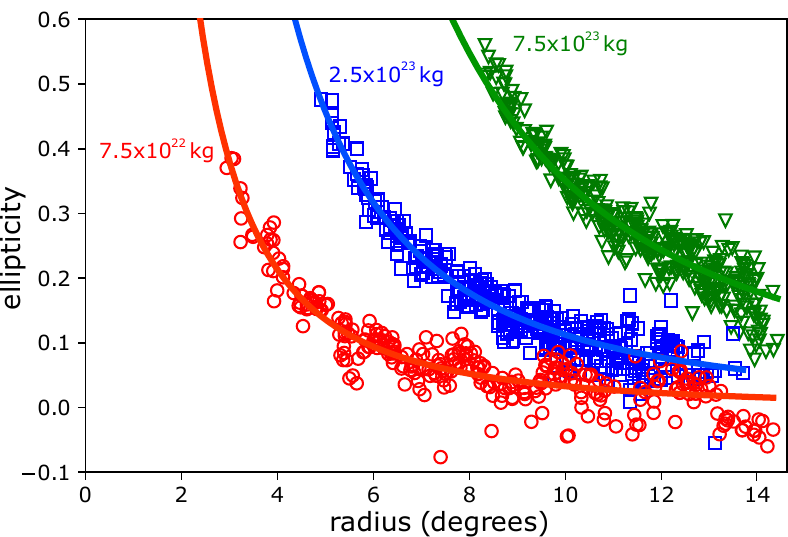}
\caption{Ellipticity as a function of angular radius for three different point-mass lenses in the weak lensing regime. The fit with Equation~\eqref{eqn:powsq} determines the Einstein radius. These plots show data up to $r_{\kappa=0.5}$.}
\label{fig:multiweak}
\end{center}
\end{figure}

Table~\ref{tab:weakmass} shows weak lensing measurements using acrylic lenses that simulate various masses. Figure~\ref{fig:multiweak} shows examples of ellipticity measurements as a function of radius. The minimum distance to the center of the lens is decreased to $r_{\kappa<0.5}$ in order to have a sufficient number of ellipses when using the largest simulated mass. These measurements show good agreement across the range of simulated masses used, demonstrating that the same analysis techniques used with images of true weak gravitational lensing can be used to make an accurate determination of the simulated mass of these lenses.

\begin{table}[ht]
\begin{ruledtabular}
\caption{Measurements of weak gravitational lensing using acrylic lenses that simulate various point-mass gravitational lenses. Ellipses with $r_{\kappa<0.5}$ are used. Values in parenthesis are one standard error due to the statistical uncertainty from the distance measurements and the fit that determines $\theta_E$.}
\begin{tabular}{cccccc}
$D_S$ (m) & $D_L$ (m) & $D_{LS}$ (m) & $\theta_E$ (degrees) & measured mass (kg) & expected mass (kg) \\
\hline
0.410 & 0.140 & 0.270 & 1.81(2) & 7.6(2)$\times 10^{22}$ & 7.5$\times 10^{22}$\\
0.410 & 0.140 & 0.270 & 3.38(2) & 2.53(6)$\times 10^{23}$ & 2.5$\times 10^{23}$\\
0.410 & 0.140 & 0.270 & 5.6(1) & 7.66(19)$\times 10^{23}$ & 7.5$\times 10^{23}$
\end{tabular}

\label{tab:weakmass}
\end{ruledtabular}
\end{table}

\subsection{Gravitational Microlensing}

Gravitational Microlensing occurs when the Einstein radius is too small to resolve individual images. According to the quadratic form of Equation~\eqref{quad}, gravitational lensing results in two solutions for the position of the image of the source object, or an Einstein ring if $\beta = 0$. If the mass of the lensing object is very small, then the Einstein angle is very small and the image of the source object cannot be resolved. The Hubble and James Webb Space Telescopes, for example, have angular resolutions on the order of 0.1 arcseconds, leaving them and other modern telescopes unable to resolve the micro-arcseconds of separation that give this type of gravitational lensing its name. Even though the images of the source object can't be resolved, the increase in brightness of the lensed image can be measured. These effects are difficult to detect in a stationary system because the unresolved lensed images cannot easily be distinguished from the unlensed source. Therefore, microlensing observations are limited to situations where the lens and source object undergo relative motion. Since closer objects have higher angular speeds on the sky than far away objects, the observer typically sees the lensing object moving in front of a distant, more stationary background source object.

\begin{figure}[ht]
\begin{center}
\includegraphics[width=3.5in]{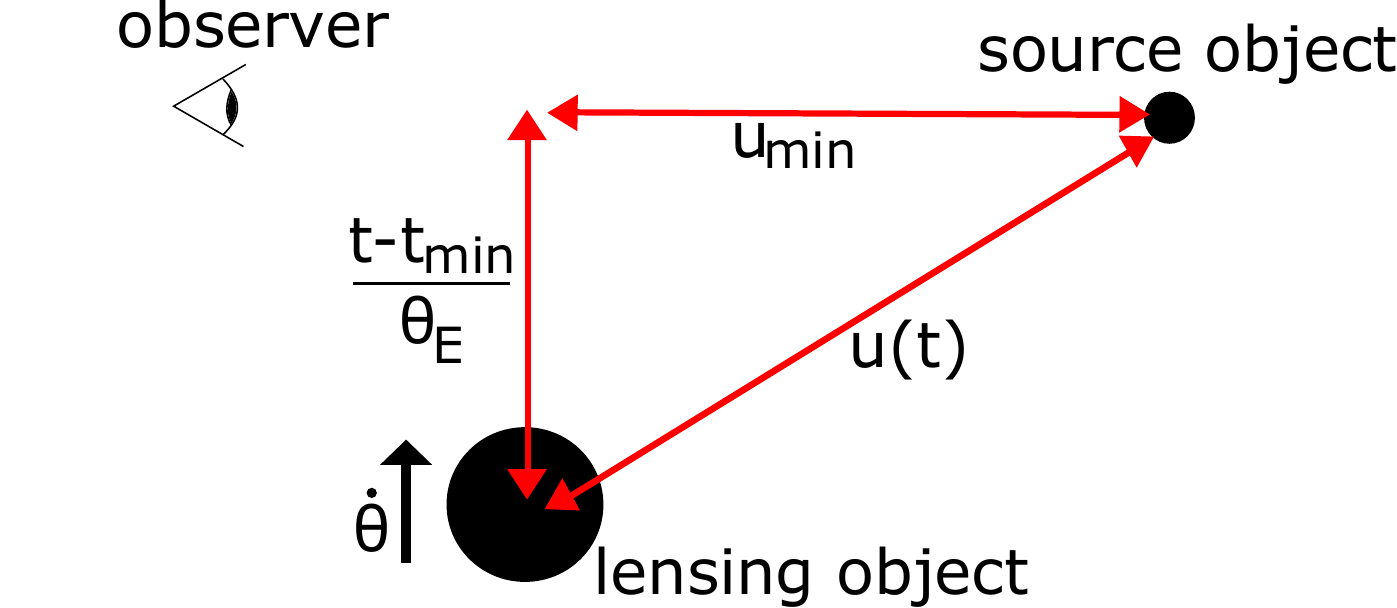}
\caption{A schematic of gravitational microlensing. The lensing object moves through the line of sight between the observer and the source object. At time $t_{\mbox{min}}$ the distance between the source and lensing objects is at a minimum $u_{\mbox{min}}$ and the brightness viewed by the observer is at a maximum. }
\label{fig:micro}
\end{center}
\end{figure}

Figure~\ref{fig:micro} shows a schematic of a gravitational microlensing system. The lensing object travels between the observer and source object at angular velocity $\dot{\theta}$. The time-dependent angular separation between the source and lensing object, normalized by the Einstein radius ($\theta_E$) is labelled $u(t)$. The magnitude of $u(t)$ reaches a minimum ($u_{\mbox{min}}$) at time $t=t_{\mbox{min}}$, which is when the combined brightness of the source object images is at a maximum. The distance traversed by the lens, relative to $u_{\mbox{min}}$, is simply the velocity of the lensing object multiplied by the time relative to $t_{\mbox{min}}$, then normalized by the Einstein radius:

\begin{equation}
\frac{\dot{\theta}(t-t_{\mbox{min}})}{\theta_E} = \frac{(t-t_{\mbox{min}})}{t_{E}},
\end{equation}

\noindent where

\begin{equation}
t_E = \theta_E/\dot{\theta}
\label{eqn:te}
\end{equation}

\noindent is the characteristic timescale of the change in measured brightness. The angular separation between the lens and source object can then be determined using the Pythagorean Theorem:

\begin{equation}
u(t) = \sqrt{u_{\mbox{min}}^2+\bigg(\frac{t-t_{\mbox{min}}}{t_{E}}\bigg)^2}.
\label{eqn:micro1}
\end{equation}

\begin{figure}[ht]
\begin{center}
\includegraphics[width=3.5in]{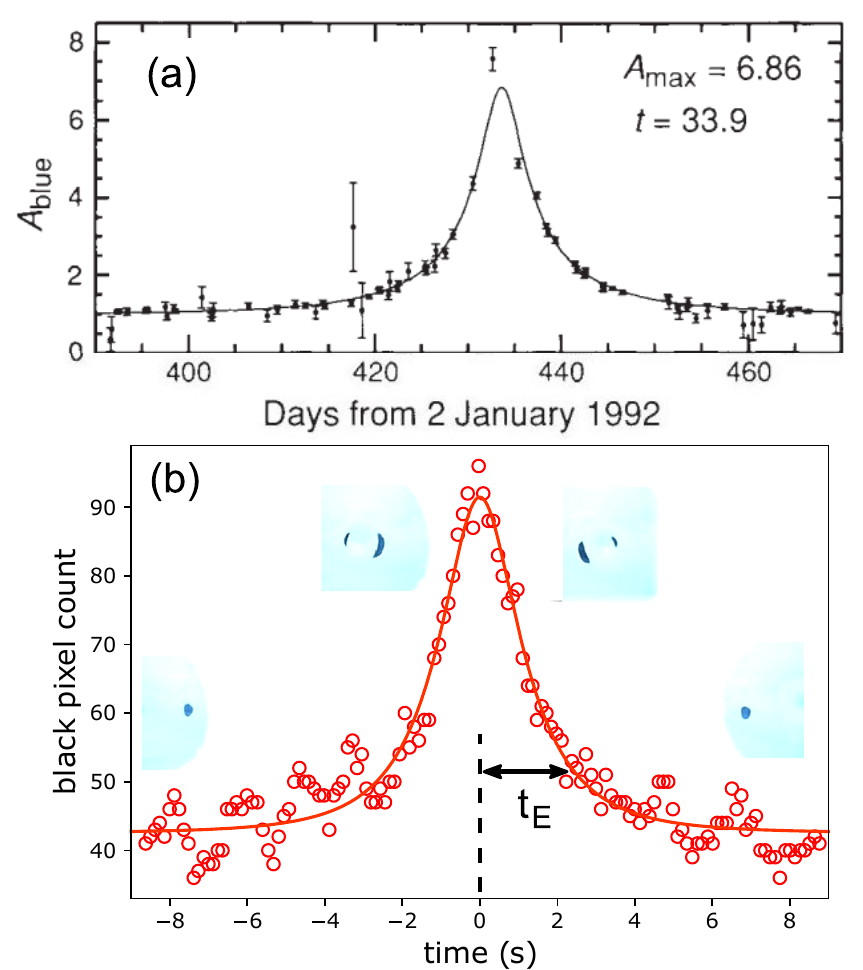}
\caption{(a) An example of changes in image brightness during a microlensing event due to a distant star~\cite{Micro}. Reprinted with permission from Springer Nature. The timescale of the change determines the Einstein radius. (b) Changes in black pixel count from moving an acrylic lens that simulates a $7.5\times 10^{22}$ kg point-mass gravitational lens in front of a 10-mm-diameter black dot. A least-squares fit with Equation~\eqref{eqn:micro2} determines $t_E$, which is used to determine the Einstein radius with Equation~\eqref{eqn:te}. The four insets show the lensed image of the source (pre-threshold) as the lens traverses the observer’s line of sight.}
\label{fig:LightCurve}
\end{center}
\end{figure}

The observed brightness depends on the angular separation between the lens and source object according to~\cite{microamp}

\begin{equation}
S(t) = S_{0}\frac{u^2(t)+2}{u(t)\sqrt{u^2(t)+4}},
\label{eqn:micro2}
\end{equation}

\noindent where $S(t)$ is the time-dependent brightness and $S_0$ is the brightness of the unlensed source object. Note that $S(t)$ never decreases below $S_0$, indicating that gravitational microlensing always results in an increase in brightness. Researchers fit the brightness measurements over time using Equation~\eqref{eqn:micro2} to determine $t_E$, then measure or model the lens’s angular velocity, calculate the Einstein radius from Equation~\eqref{eqn:te}, and determine the mass using Equation~\eqref{eqn:mass}. Figure~\ref{fig:LightCurve}(a) shows an example of brightness measurements from the gravitational microlensing of a distant star. 

This simulation of gravitational microlensing uses a webcam to view the lensed image of a 10-mm-diameter black dot on a sheet of white paper as the acrylic lens passes between them. Python, with OpenCV, thresholds the image into either black or white pixels, then counts the number of black pixels in each frame from the webcam. The number of black pixels in each frame is a useful proxy for the brightness that the observer would measure in true gravitational microlensing. 

\begin{figure}[ht]
\begin{center}
\includegraphics[width=5in]{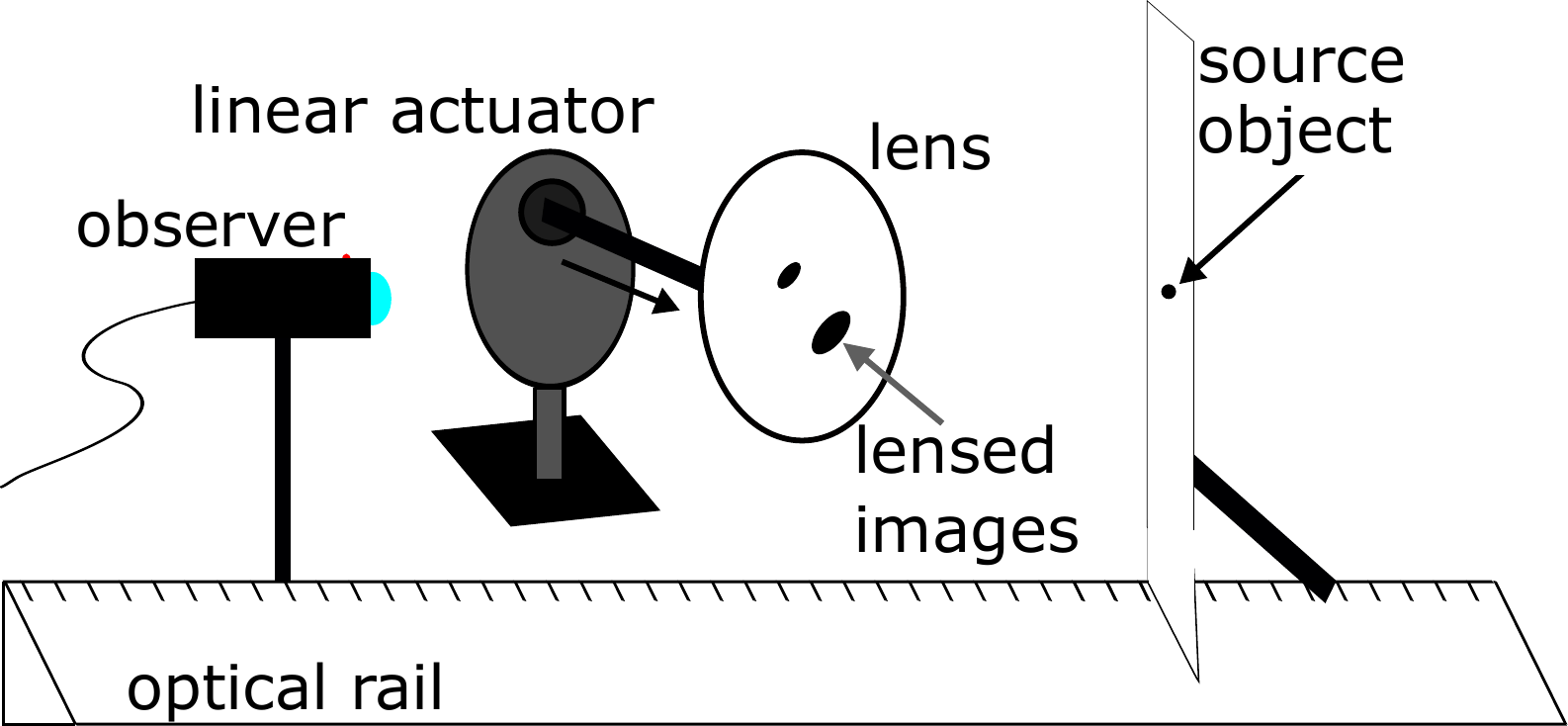}
\caption{Schematic diagram of the apparatus used to simulate gravitational microlensing. The linear actuator moves the lens out of the page. The image of the dot is magnified and duplicated as the lens passes between the observer and the source image.}
\label{fig:microapp}
\end{center}
\end{figure}

A linear actuator (Hydroworks 8489072) moves the lens perpendicular to the line between the observer and source object, as shown schematically in Figure~\ref{fig:microapp}. The impact parameter of the passing lens is chosen to be large enough that the image of the source object is never a complete Einstein ring, which would appear as a plateau in the light curve. The number of black pixels increases as the lensed image is magnified and duplicated as the lens approaches $u_{\mbox{min}}$, then decreases as the lens moves away. Figure~\ref{fig:LightCurve}(b) shows the black-pixel count as a function of time and a least-squares fit using Equation~\eqref{eqn:micro2} is used to solve for $S_0, u_{\mbox{min}}, t_{\mbox{min}}$, and $t_E$. Measured values of $t_E$ and $\dot{\theta}$ (calculated from the measured linear speed of the actuator and distance to the lens) determine the Einstein radius, then the simulated mass of the acrylic lens is calculated using Equation~\eqref{eqn:mass}. Table~\ref{tab:micro} shows measurements of the simulated mass at various positions on the optical rail. The weighted average of these measurements is 7.1(4)$\times 10^{22}$ kg, which is in good agreement with the expected simulated mass of $7.5 \times 10^{22}$ kg. 

\begin{table}[ht]
\begin{ruledtabular}
\caption{Measurements of gravitational microlensing using an acrylic lens that simulates a $7.5 \times 10^{22}$ kg point-mass gravitational lens. The fit in Figure~\ref{fig:LightCurve}(b) determines $t_E$ and its uncertainty. Values in parenthesis are one standard error due to the statistical uncertainty from the distance and speed measurements, and the fit that determines $t_E$.}
\begin{tabular}{ccccccc}
$D_S$ (m) &$ D_L$ (m) &$ D_{LS}$ (m) & $t_E$ (s) & $\dot{\theta}$ (degrees/s) & $\theta_E$ (degrees) & mass (kg) \\
\hline
0.750 & 0.290 & 0.460 & 2.15(8) & 0.61(2) & 1.31(7) & 8.4(9)$\times 10^{22}$ \\
0.615 & 0.235 & 0.380 & 1.98(12) & 0.76(3) & 1.52(10) & 9.0(1.2)$\times 10^{22}$ \\
0.615 & 0.335 & 0.280 & 1.37(3) & 0.665(17) & 0.91(3) & 6.3(5)$\times 10^{22}$ \\
0.405 & 0.125 & 0.280 & 1.13(6) & 0.178(6)& 2.0(1) & 7.6(1.0)$\times 10^{22}$ \\
\end{tabular}
\label{tab:micro}
\end{ruledtabular}
\end{table}

\begin{figure}[ht]
\begin{center}
\includegraphics[width=3.5in]{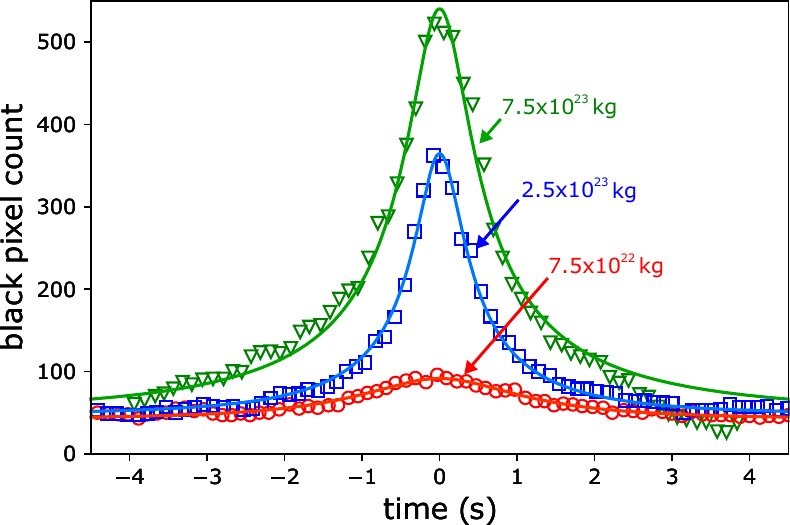}
\caption{A Python script counts the number of black pixels in each frame from a webcam as the acrylic lens passes between the observer and source object.  Measurements are taken for three different point-mass lenses in the gravitational microlensing regime. A least-squares fit with Equation~\eqref{eqn:micro2} determines $t_E$, which is used to determine the Einstein radius with Equation~\eqref{eqn:te}.}
\label{fig:twomicro}
\end{center}
\end{figure}

Table~\ref{tab:micro1} shows gravitational microlensing measurements using lenses that simulate various masses. When using the 7.5 $\times 10^{23}$ kg lens, the image of the source object is magnified and duplicated over the entire diameter of the lens as it passes in front of the source object, so the number of black pixels never reaches a constant, unlensed value, as shown in Figure~\ref{fig:twomicro}.  The least squares fit with Equation~\eqref{eqn:micro2} cannot determine $S_0$, so the value is fixed at the average of the two lighter lenses (46(2) black pixels). The unlensed brightness and width of the curve are strongly correlated, thereby increasing the uncertainty in the mass measurement of this lens. All three mass measurements are in reasonable agreement with the expected value of the simulated mass based on the machined curvature of the lens. The degree of agreement between these measurements and the expected mass indicate that this simulation of microlensing is a good analogy for true gravitational microlensing.

\begin{table}[ht]
\begin{ruledtabular}
\caption{Measurements of gravitational microlensing using acrylic lenses that simulate various point-mass gravitational lenses. Values in parenthesis are one standard error. The uncertainty in the mass measurement of the 7.5 $\times 10^{23}$ kg lens includes the uncertainty in the unlensed black pixel count that could not be measured directly.}
\begin{tabular}{cccccccc}
$D_S$ (m) &$ D_L$ (m) &$ D_{LS}$ (m) & $t_E$ (s) & $\dot{\theta}$ (degrees/s) &$\theta_E$ (degrees) & measured mass (kg) & expected mass (kg) \\
\hline
0.750 & 0.290 & 0.460 & 2.15(8) &0.61(2) &1.31(7) & 8.4(9)$\times 10^{22}$ & 7.5$\times 10^{22}$ \\
0.750 & 0.290 & 0.460 & 2.40(7) & 0.89(2)&2.15(8) & 2.24(17)$\times 10^{23}$ &2.5 $\times 10^{23}$ \\
0.750 & 0.290 & 0.460 & 4.9(1) & 0.89(2)&4.34(15) & 9(1) $\times 10^{23}$ &7.5 $\times 10^{23}$ \\

\end{tabular}
\label{tab:micro1}
\end{ruledtabular}
\end{table}

\section{Conclusion}

Astrophysics is difficult to incorporate into the undergrad lab since much of the work relies on observations of celestial objects with telescopes and requires specialized software to stack and process the images. This experiment allows students to control the parameters of the lensing observation such as distance scales, masses, and source objects, see the resulting lensed images in real-time, and analyze those images using freely available software. These experiments simulate all three regimes of gravitaitonal lensing, corresponding to increasing levels of complexity.

The strong gravitational lensing experiment is well suited to a second-year laboratory course, as it requires prior exposure to optics and basic astronomy to appreciate the underlying analogy. In our labs, students move the lens along the optical rail and determine how the position affects the mass measurement, leading them to discover the small angle approximation and identify the region of the optical rail where the approximation remains valid.  Students then measure the emulated mass of six different lenses, fit the resulting data with a line and evaluate whether the slope differs significantly from unity, while considering possible systematic effects and choice of fitting model ($mx+b$ versus $mx$).

The weak lensing and microlensing experiments are better suited to upper-year laboratory courses.  The microlensing data analysis requires a nonlinear fit and considerable patience to achieve adequate lighting and image contrast. In our labs, students take measurements with different lenses in the microlensing regime and discover the limitation of the least-squares fit with the heaviest lens. The data analysis of the weak lensing experiment is the most challenging activity, though the ellipse selection steps provide opportunity for critical thinking and iterative analysis. Junior undergraduate students are unlikely to achieve accurate results across all three lensing regimes without significant scaffolding of the analysis procedure.

The scope of this simulated lensing experiment is limited by modelling the lensing body as a $\sim10^{23}$ kg point-mass, although no confirmed lunar-mass black holes have been observed.  Gravitational lenses are often galaxies or clusters of galaxies and better described by an isothermal mass profile, which can also be simulated with an acrylic lens~\cite{mainlens}. An isothermal lens could be used to test assumptions about the expected trend in the weak lensing data, since the ellipticity should decrease as $1/r$ in an isothermal mass profile rather than $1/r^2$ for the point mass lens. 

The techniques used to analyze the images from these experiments are analogous to modern techniques used to analyze observations of true gravitational lensing. This simulation provides students with experience in model fitting, image analysis, and interpretation of observational data within an accessible laboratory setting.  These experiments offer a practical introduction to modern observational astronomy while remaining feasible for undergraduate instruction.

\section{Supplement}

Please click on this link to access the supplementary material, which includes ImageJ macros for image analysis, STEP files for the lenses, and Python code for lens generation, data collection, and analysis. Print readers can see the supplementary material at [DOI to be inserted by AIPP].

\section{Acknowledgements}

 I would like to thank the reviewers and editorial staff for their valuable feedback, which greatly improved this work.  Thank you to the McMaster University Engineering Research Machine Shop for fabricating the lenses and to Dr. Laura Parker for helpful conversations.

\end{document}